\documentclass[doublecol]{epl2}
\usepackage{color}
\usepackage{bm}
\usepackage{latexsym}
\usepackage{amssymb}
\usepackage{amsmath}
\usepackage{graphicx}
\usepackage{ulem}
\begin{document}

\newcommand{\half}{\mbox{\small $\frac{1}{2}$}}
\newcommand{\cc}{\mbox{c.c.}}
\newcommand{\abs}[1]{\lvert #1 \rvert}
\newcommand{\avg}[1]{\langle #1 \rangle}
\newcommand{\bra}[1]{\left\langle #1\right|}
\newcommand{\ket}[1]{\left|#1\right\rangle }
\newcommand{\edit}[1]{\textcolor{red}{#1}}
\newcommand{\nin}{n_{in}}
\renewcommand{\etal}{\textit{et al.}}

%%%%%%%%%%%%%%%%%%%%%%%%%%%%%%%%%%%%%%%%%%
\title{Two interacting particles in a random potential}
%%%%%%%%%%%%%%%%%%%%%%%%%%%%%%%%%%%%%%%%%%

%\author{Dmitry O. Krimer${}^{1,2}$ , Ramaz Khomeriki${}^{1,3}$  and Sergej Flach${}^{1}$ }
%\institute{${\ }^1$Max Planck Institute for the Physics of Complex Systems, N\"othnitzer Stra\ss e 38, D-01187 Dresden, Germany
%${\ }^2$Theoretische Physik, Universit\"at T\"ubingen, Auf der Morgenstelle 14,
%72076 T\"ubingen, Germany
%${\ }^3$Physics Department, Tbilisi State University, Chavchavadze 3, 0128 Tbilisi, Georgia}

\author{Dmitry O. Krimer\inst{1,2}\thanks{E-mail: \email{dmitry.krimer@gmail.com}}\and Ramaz Khomeriki\inst{1,3} \and Sergej Flach\inst{1}}
\shortauthor{Krimer, Khomeriki and Flach}
\institute{                    
  \inst{1}Max Planck Institute for the Physics of Complex Systems, N\"othnitzer Stra\ss e 38, D-01187 Dresden, Germany\\
  \inst{2} Theoretische Physik, Universit\"at T\"ubingen, Auf der Morgenstelle 14,
72076 T\"ubingen, Germany\\
 \inst{3}Physics Department, Tbilisi State University, Chavchavadze 3, 0128 Tbilisi, Georgia
}


\abstract{We study the scaling of the localization length of two interacting particles in
a one-dimensional random lattice with the single particle localization length.
We obtain several regimes, among them one interesting weak Fock space disorder regime.
In this regime we derive a weak logarithmic scaling law.
Numerical data support the absence of any strong enhancement of the
two particle localization length.}

\maketitle

%%%%%%%%%%%%%%%%%%%%%%%%%%%%%%%%%%%%%%%%%%
\section{Introduction}
%%%%%%%%%%%%%%%%%%%%%%%%%%%%%%%%%%%%%%%%%%

Quantum single particle dynamics in
one-dimensional
disordered lattices with uncorrelated random onsite energies
exhibits Anderson localization \cite{PWA58}.
The asymptotic spatial decay of an eigenvector is exponential and given by
$A_{l}^{(\nu)} \sim {\rm e}^{-l/\xi_1^{\nu}}$, where $\xi_1^\nu$
is the localization length of an eigenmode $\nu$ with the eigenvalue
$\lambda_\nu$, and the integer $l$ counts the lattice site (see also e.g. \cite{KRAMER}).
The localization length is bounded from above.
%The most
%extended modes correspond to the bandwidth center with
%$\xi_1(\lambda=0,W) \approx 100/W^2$ for $W\leq4$. This result was
%obtained both in the framework of second order perturbation theory
%in the limit of weak disorder and using the standard
%transfer-matrix approach (see e.g. the review \cite{KRAMER}).

The interplay of disorder and interaction
of two interacting particles (TIP) in a random
one-dimensional potential stirred was considered by Shepelyansky (Sh94) \cite{Shep_94}.
The conclusion was that
two particles might propagate coherently over distances
much larger than the single particle
localization length $\xi_1$, if both particles are
launched within a distance of $\xi_1$ from each other. Sh94 used an analogy between
the two-particle eigenvalue problem and that of
banded random matrices, and made an assumption about the scaling properties of overlap integrals
which connect different noninteracting Fock eigenstates in the presence of
interaction. He finally concluded
that in the weak disorder limit $\xi_1 \rightarrow \infty$ the two-particle localization length
$\xi_2$ will scale with $\xi_1$ as $\xi_2 \varpropto \xi_1^2U^2$,
where $U$ is the interaction strength \cite{Shep_94}. This result was further supported by
Imry (Im95)
in \cite{Imry_95}, where a Thouless-type scaling
argument was replacing the banded random matrix analogy.
Therefore, two interacting particles were predicted to explore
a much larger space than noninteracting particles.
%It is worth noting, that a randomness of each
%term in the sum for the overlap integral of single particle wave
%functions, was the key assumption made in \cite{Shep_94,Imry_95}.
Numerical calculations by Frahm et al (FR)
\cite{frahm95} concluded that the scaling is probably weaker, namely
$\xi_2 \varpropto \xi_1^{1.65}$, and raised doubts about the previously assumed
scaling properties of overlap integrals.
% showed that the
%Subsequently, taking into account a more accurate distribution for
%the overlap of single particle functions, the localization length
%$\xi_2$ was found to scale with $\xi_1$ with a smaller exponent,
%namely $\xi_2 \varpropto \xi_1^{1.65}$ \cite{frahm95}.
Using
a Green function method adapted to the problem, a new scaling
relation at the center of the band, $\xi_2=\xi_1+c \xi_1^2 |U|/V$,
was derived in \cite{Oppen_96} ($c\approx 0.11$ for bosons). In
particular, this implies that the enhancement effect will set in for
weaker interactions than previously predicted. Later on, it was
argued that the enhancement effect is probably due to finite-size
effects and it should completely vanish for an infinite system
\cite{Screib_97}. Simulating the time dependent Schr\"odinger
equation for two interacting particles \cite{Waintal99}, it was
argued that the dynamics is characterized by two time scales, $t_1$ and
$t_2$, set by, respectively, two localization lenghts, $\xi_1$ and
$\xi_2$.
%In particular, for $t<t_1$ ($t_1<t<t_2$) the center of
%mass was shown to spread ballistically (subdiffusively).
%This numerical path is however not appropriate for a precise measurement
%of the scaling relation under debate.
%method is, however, not appropriate for an estimation the scaling
%realtion $\xi_2/\xi_1$.
Recently, two of us studied statistical
properties of the overlap integrals perturbatively and numerically for weak
disorder \cite{kf10}. These results contradict previous assumptions of Sh94 and Im95
\cite{Shep_94,Imry_95}, and if used within the previously applied theoretical schemes,
predict a much weaker interaction induced increase of the localization length than
previously discussed.  Despite a number of studies, the problem of two interacting
particles in a random potential remains therefore a completely open problem.
At the same time this seemingly academic case can be both addressed by current techniques
with ultracold interacting atoms \cite{wtl06}, and is of fundamental importance for tackling
the much more complicated case of many interacting particles in random potentials.

In the present work we first show that a nonperturbative strong localization length enhancement
can be expected only in a regime of very weak disorder, with upper bounds on the
disorder strength. This regime was not fully accessed in previous numerical scaling studies.
We then obtain upper bounds on the strength of the expected enhancement
effect using correct scaling properties of overlap integrals.
We then perform direct
numerical measurements of the two-particle localization length in the perturbative
and nonperturbative regimes,  by solving the corresponding eigenvalue problem
with subsequent
averaging over many disorder realizations. Finally we formulate a set of open issues
which have to be addressed in the future.

%%%%%%%%%%%%%%%%%%%%%%%%%%%%%%%%%%%%%%%%%%
\section{Model}
%%%%%%%%%%%%%%%%%%%%%%%%%%%%%%%%%%%%%%%%%%

We consider the Bose-Hubbard Hamiltonian with disorder
\begin{eqnarray}
&&{\cal \hat H}\equiv{\cal \hat H}_0+{\cal \hat H}_{int}\;,\;
{\cal \hat H}_{int} = \sum\limits_l\left[ \frac{U}{2}\hat a_{l}^+\hat a_l^+\hat a_{l}\hat a_l\right]\;,
\\\nonumber
&&{\cal \hat H}_0 = \sum\limits_l\left[\epsilon_l\hat a_{l}^+\hat a_l+V \left( \hat a_{l+1}^+\hat a_l+\hat
a_{l}^+\hat a_{l+1}\right) \right], \label{eq_Hamilt}
\end{eqnarray}
which consists of non-interacting and interacting parts, ${\cal
\hat H}_0$ and ${\cal \hat H}_{int}$. Here $\hat a_{l}^+$ and
$\hat a_{l}$ are standard boson creation and annihilation
operators on a lattice site $l$ and $U$ measures the interaction
strength. The random on-site energies $\epsilon_{l}$ are chosen
uniformly from the interval $[-W/2,W/2]$, with $W$ and $V$
denoting the disorder and hopping strengths, respectively.

\subsection{One particle}
In this case the interaction term does not contribute.
%play any role and
We use
the basis $|l\rangle\equiv a_{l}^+|0\rangle$ with
$l=1,\ldots ,N$ ($N$ is the number of lattice sites) .
%which are the
%eigenstates of a number operator $\hat{N}=\sum_l a_{l}^+a_{l}$.
The eigenstates (also called single particle normal modes (NM))
$|\nu\rangle=\sum_{l}^NA_{l}^{(\nu)}| l \rangle$ are defined through the eigenvectors
$A_l^{(\nu)}\sim e^{-|l|/\xi_1^\nu}$
%are the single particle
%normal modes (NM) given by
with the eigenvalue problem
\begin{equation} \lambda_{\nu} A_l^{(\nu)} = \epsilon_l
A_l^{(\nu)} +V(A_{l+1}^{(\nu)} + A_{l-1}^{(\nu)}). \label{AA}
\end{equation}
The eigenvalues $-2V-W/2 \leq \lambda \leq 2V+W/2$ fill a band
with a width $\Delta_1 = 4V+W$. The most extended NMs correspond to
the band center $\lambda=0$ with localization length
\begin{equation}\label{00}
\xi_1(\lambda=0,W) \approx 100(V^2/W^2),
\end{equation}
in the limit of weak disorder $W/V\leq4$ \cite{KRAMER}.
The average volume $L$ which an eigenstate occupies has been estimated to be about $L\approx 3\xi_1$ for weak disorder \cite{kf10}.

\subsection{Two particles}
%
%Let us start with $U=0$.
%Taking $V=1$ we consider the noninteracting case, $U=0$. The problem reduces again to
%the single particle problem and one
For $U=0$ we construct orthonormalized
two particle eigenstates as product states of single particle eigenstates in a corresponding
Fock space
\begin{equation}
\label{eigen_H0} | \mu,\nu\geq\mu \rangle=\frac{|\mu\rangle|\nu
\rangle}{\sqrt{1+\delta_{\mu,\nu}}}\;,\; {\cal \hat H}_0 |
\mu,\nu\rangle=(\lambda_\mu+\lambda_\nu)  | \mu,\nu \rangle.
\end{equation}
Then, we expand the eigenstates $| q\rangle$ of the
interacting particle problem, ${\cal \hat H} | q\rangle=\lambda_q
|q\rangle$, in systems of eigenstates for the noninteracting problem,
$|q\rangle=\sum_{\nu, \mu\le \nu}^N
\phi_{\mu\nu}^{(q)} |\mu,\nu \rangle$, where the coefficients
$\phi_{\mu\nu}^{(q)}$ satisfy the eigenvalue problem
\begin{equation}
\lambda_q\phi_{\mu\nu}^{(q)}=\lambda_{\mu\nu}\phi_{\mu\nu}^{(q)}
+2U\sum\limits_{\mu',\nu'}
\bar{I}_{\mu\nu}^{\mu'\nu'}\phi_{\mu'\nu'}^{(q)}. \label{eq1511}
\end{equation}
Here $\lambda_{\mu\nu}\equiv \lambda_{\mu}+\lambda_{\nu}$ and therefore the noninteracting case $U=0$
yields a band with width $\Delta_2=2\Delta_1$. The
coefficients $\bar{I}_{\mu\nu}^{\mu'\nu'}$ are connected with the
overlap integrals
\begin{equation}
I_{\mu\nu}^{\mu'\nu'} = \sum_{l}A_l^\mu A_l^\nu
A_l^{\mu'}A_l^{\nu'}
\label{overlapintegral}
\end{equation}
as follows:
\begin{eqnarray}
\bar{I}_{\mu\nu}^{\mu'\nu'}=\frac{I_{\mu\nu}^{\mu'\nu'}}{\sqrt{1+\delta_{\mu\nu}}\sqrt{1+\delta_{\mu'\nu'}}}.
\label{eq16}
\end{eqnarray}
The interacting case yields a single band for $U < \Delta_2$, but
two bands separated by a gap for $U > \Delta_2$. Indeed, in the
latter case two-particle bound states are renormalized out of the
main band, and are mainly consisting of two particles occupying
the same site \cite{eilbeck}. Therefore, remaining band is due to
states where the two particles can be anywhere but not on the same
site. This is simply the limit of two noninteracting spinless
fermions. The localization length of these two noninteracting
fermions is of the same order as the single particle localization
length. The localization length in the bound state band is even
smaller, since the effective disorder strength in this band
becomes $2W$, but the effective hopping is strongly suppressed.

For numerical purposes we expand the two particle eigenstates
$|q\rangle$ in the local basis
\begin{equation}
|q\rangle=\sum_{m, l\le m}^N{\cal L}_{l,m}^{(q)}| l,m \rangle,
\quad  |l,m \rangle\equiv
\frac{a_{l}^+a_{m}^+|0\rangle}{\sqrt{1+\delta_{lm}}},
\label{eq_plh}
\end{equation}
where ${\cal L}_{l,m}^{(q)}=\langle l,m | q\rangle$ are the
normalized eigenvectors. They satisfy
\begin{eqnarray}
\label{eq_proj} \phi_{\mu\nu}= \sum_{m, l\le m}^N
\dfrac{A_m^{(\mu)}A_l^{(\nu)}+A_l^{(\mu)}A_m^{(\nu)}}{\sqrt{1+\delta_{lm}}
\cdot\sqrt{1+\delta_{\mu \nu}}}\cdot{\cal L}_{l,m}^{(q)}\;.
\end{eqnarray}

We will numerically compute the probability density
function (PDF) of the number of particles in direct space $p_l=\langle q | \hat a_{l}^+\hat a_l |
q\rangle/2$, which is given by
\begin{equation}
p_l^{(q)}=\dfrac{1}{2}\left(\sum_{k, l\le k}^N{\cal
L}_{l,k}^{(q)2}+ \sum_{m, l\ge m}^N {\cal L}_{m,l}^{(q)2}\right).
\label{eq_pl}
\end{equation}

\section{On scales}

Since a single particle eigenstate occupies a volume $L$, there
are of the order of $L^2$ two particle eigenstates which are
residing in the same volume for $U=0$. The overlap integrals built
among these  $L^2$ Fock states are nonzero (more precisely not
exponentially weak) and define the connectivity in the Fock space
for nonzero $U$. The average eigenenergy spacing $d$ of these
connected Fock states is $d=\Delta_2 / L^2$. It therefore defines
an effective energy mismatch - i.e. an effective disorder strength
$\bar{W}\equiv d$ - in the Fock space. The effective hopping
strength follows from (\ref{eq1511}) and is given by $\bar{V} = 2
U \langle I \rangle$. Here $ \langle I \rangle$ is an average
overlap integral among all connected Fock states \cite{kf10}.

In analogy with eq. (\ref{00}) we can therefore obtain a localization length in Fock space
for weak Fock space disorder $\bar{W} \lesssim 4 \bar{V}$,
which in real space is a measure in units of the single particle localization length:
\begin{equation}
\frac{\xi_2}{\xi_1} \approx 100 \frac{\bar{V}^2}{\bar{W}^2} =
400 \frac{U^2}{\Delta_2^2} \langle I \rangle^2 L^4\;.
\label{tpll}
\end{equation}
For strong Fock space disorder $\bar{W} \gg \bar{V}$ the volume $L\approx 1$,
and two interacting particles are localized in the same way, therefore
$\xi_2 \approx \xi_1$ in this case.

\subsection{Bounds on the weak Fock space disorder regime}

Let us now address the question whether we can enter the weak Fock
space disorder regime for strong single particle disorder $W \gg
V$. This seems possible at a first glance since we can increase
the value of $\bar{V}$ by increasing $U$. However, in this limit
$\langle I \rangle \sim V^2/W^2$. Therefore the needed
interaction strength is $U \sim W^3/V^2$, since $\Delta_2 \sim W$.
But an increase of the interaction strength beyond the band width
$\Delta_2$ leads to the separation of the energy spectrum into two
bands - a bound state band with strongly localized particle pairs
\cite{eilbeck}, and a noninteracting spinless fermion band which
has no localization length increase as compared to the single
particle case. The two conditions $U \lesssim W$ and $U \gtrsim W^3/V^2$ imply
that $W  \lesssim V$ is needed, which means that the single particle case
must be in the regime of weak localization. Therefore $U  \lesssim V$ is
an upper bound for entering the weak Fock space disorder regime.

Lowering $U$ further we will however again leave this regime and enter the perturbative one,
which is again characterized by strong disorder in Fock space. Indeed, the energy renormalization
of a given Fock state follows from (\ref{eq1511}) and is given by $2U I_0$ where $I_0$ is an average
overlap integral of a Fock state with itself. Due to orthonormality of the single particle
eigenfunctions it follows $I_0 \approx 1/L$.
The perturbative regime holds as long as $UI_0  \lesssim d$.
Inside the perturbative regime a Fock state is still a good approximation to an exact
eigenstate, and therefore the two particle localization length is of the order of
the single particle one. Therefore, the nonperturbative weak Fock space
disorder regime is accessed for $ \Delta_2/L  \lesssim U  \lesssim V$.

For any practical purposes we seek a strong enough interaction strength $U$, and  this
requires $U \approx V$ and $W < V$. In order to obtain any relevant scaling results
upon variation of $W$ one needs therefore to lower $W$ significantly further such that $W\ll V$.

\subsection{Overlap integrals revisited}

Sh94 and Im95 estimated the average overlap integral $\langle I
\rangle_{SI} \sim L^{-3/2}$ \cite{Shep_94,Imry_95} inside the weak
Fock space disorder regime. This result is obtained in the
following way. A single particle eigenstate occupies a volume
$L\gg 1$. Due to normalization it follows $|A_{l}^{(\nu)}| \sim
L^{-1/2}$. The crucial point was to assume that all terms inside
one localization volume in the sum (\ref{overlapintegral}) have
uncorrelated signs. This leads to the above estimate. However, in
the limit of weak disorder and large localization length, the
single particle eigenvectors inside a localization volume will
appear similar to plane waves, with appreciable phase correlations
between different sites, and also between different eigenstates.
Some numerical studies by R\"omer et al (R99) \cite{rsv99} {even
concluded that $\langle I \rangle_{R} \sim L^{-2}$. This result
essentially corresponds to the assumption that the eigenvectors
are exact plane wave states inside a localization volume. It is
this small difference in the exponent which separates a possible
existing strong enhancement of the localization length from no
effect at all.

In a recent work two of us performed a perturbation approach at the weak disorder limit
and obtained that strong phase correlations will certainly modify the prediction of Sh94,Im95.
At the same time corrections to the result of R99 are significant. As a final result
we obtain $\langle I \rangle_{SI} \sim -\ln{(L)} L^{-2}$ \cite{kf10} - logarithmic corrections
to the prediction of R\"omer et al. It is well-known that logarithmic corrections are rather
resistent to numerical verifications, if no special trick or technique is used.
Therefore, our numerical tests in a limited interval of $W$ lead only
to the clear result that the prediction of Sh94,Im95 is incorrect, and if
$\langle I \rangle \sim 1/L^z$ is assumed, then $z \approx 1.7$.
They were not sensitive to
distinguish between this power law and a possible asymptotic $\langle I \rangle \sim -\ln{(L)}L^{-2}$
logarithmic law.

\subsection{Scaling of the localization length}

Combining the above predictions on the overlap integral scaling and the
localization length scaling (\ref{tpll}) we arrive at the following results in the
weak Fock space disorder regime. Here we set $\Delta_2 = 8V$, take $W < 4V$ such that
(\ref{00}) holds. Then Sh94 and Im95 predict
$\xi_2/\xi_1 \sim (U/V)^2 \xi_1$ as derived using different methods in the original
papers \cite{Shep_94,Imry_95}. According to R99 the whole effect is simply
$\xi_2/\xi_1 \sim (U/V)^2$, i.e. no enhancement at all.
Finally, our analytical estimate for the overlap integrals yields
\begin{equation}
\frac{\xi_2}{\xi_1} \sim
(\ln{\xi_1})^2\left(\frac{U}{V}\right)^2\;. \label{ourscaling}
\end{equation}
Note that the numerically estimated overlap integral dependence on $L$ results in
$\xi_2/\xi_1 \sim (U/V)^2 \xi_1^{0.6}$.

\begin{figure}
%\hspace*{-1.6cm}
\includegraphics[angle=0,width=0.8\columnwidth]{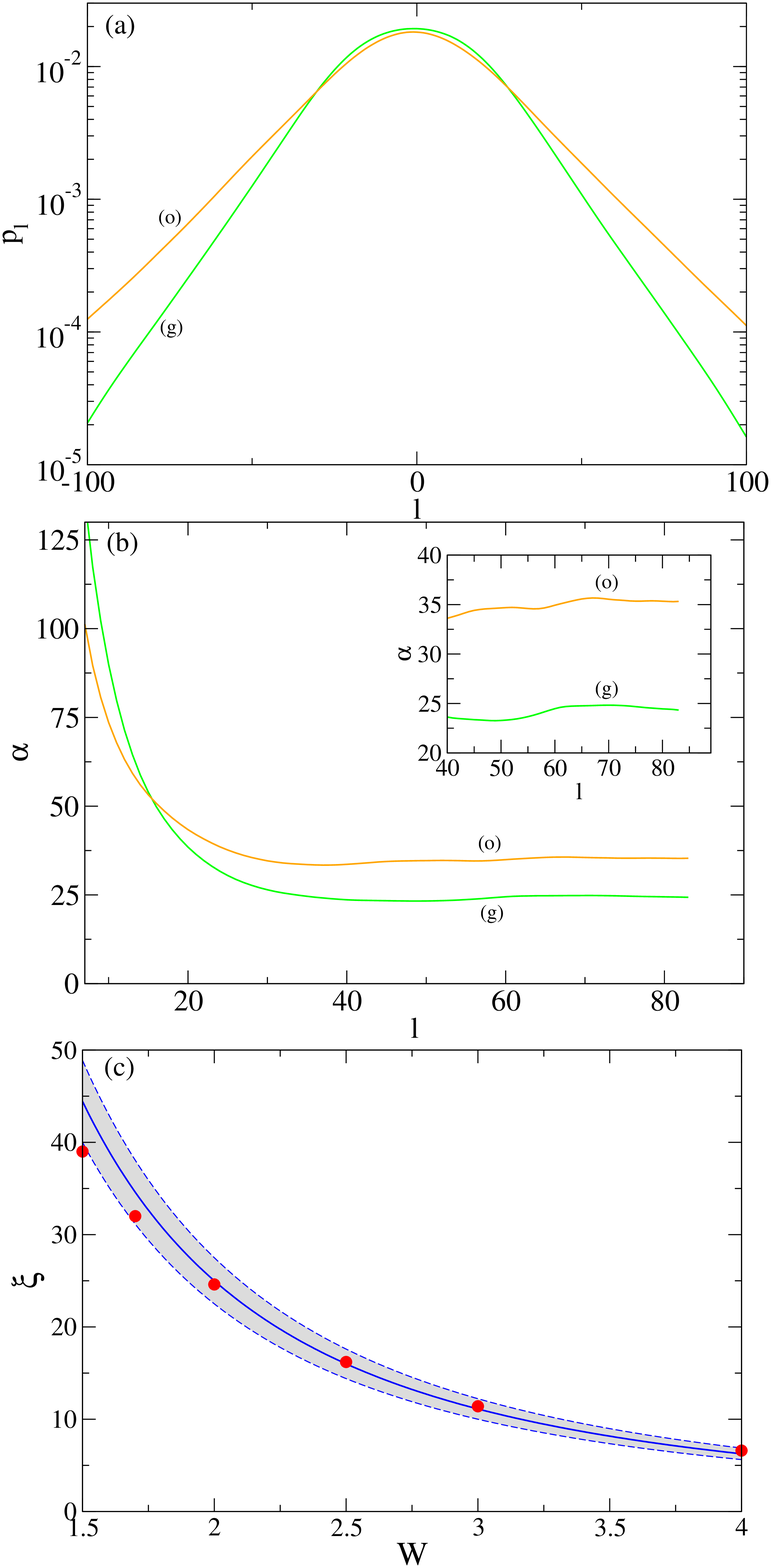}
\caption{(a) Smoothed average probability distribution function
$\langle p_l \rangle$ versus lattice site $l$ in lin-log scale for
$W=2$, $U=0$ [(g), green curve] and $W=2$, $U=0.2$ [(o), orange curve]; (b)
the corresponding quantity $\alpha$ (see text) versus $l$, with a
zoom of the interval with saturated values of $\alpha$ (inset) ;
(c) the two-particle localization length $\xi_2$ versus $W$ for
the noninteracting case, U=0 (red circles). Blue solid line:
$\xi_1=100/W^2$. Dashed lines: maximal admissible error of $10\%$
from the analytical formula. Gray area corresponds to the
admissible values.} \label{fig_W2p0_joint}
\end{figure}
\begin{figure}
\includegraphics[angle=0,width=0.8\columnwidth]{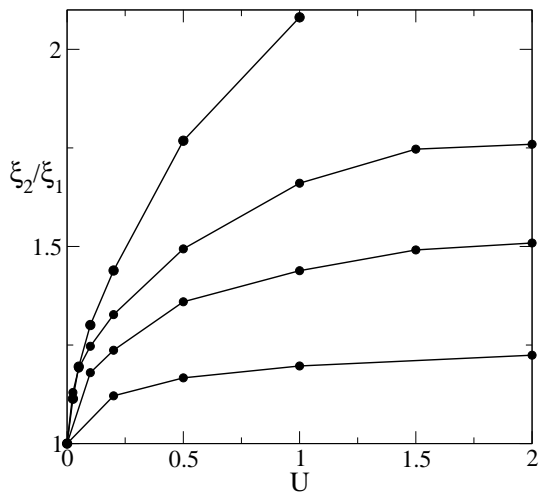}
\caption{The ratio $\xi_2/\xi_1$ versus interaction constant
$U$ for different values of disorder $W=2,2.5,3,4$  (from top to
bottom)} \label{fig_xi2_vs_U}
\end{figure}
\begin{figure}
\includegraphics[angle=0,width=0.8\columnwidth]{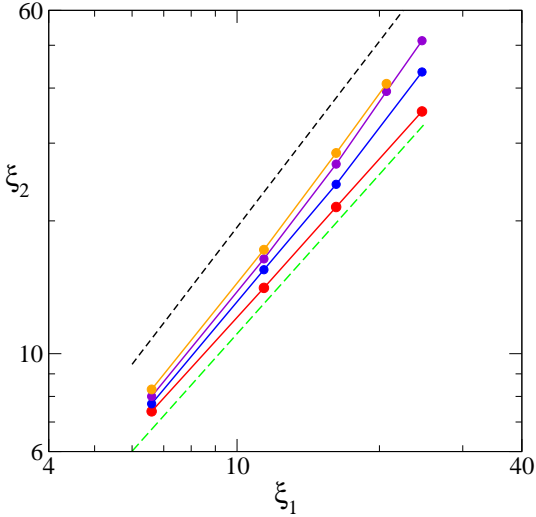}
\caption{The two-particle localization length $\xi_2$ versus
one particle localization length $\xi_1$ for $U=1.5,1.0,0.5,0.2$
(from top to bottom) in log-log scale. Dashed straight lines are
power laws $\xi_1^\alpha$ with the exponents $\alpha=1.4$ (upper
line) and $\alpha=1.3$ (lower line). The size
of a chain is $N=234$.} \label{fig_xi2_xi0}
\end{figure}

\begin{figure}
\hspace*{-1cm}
\includegraphics[angle=0,width=1.1\columnwidth]{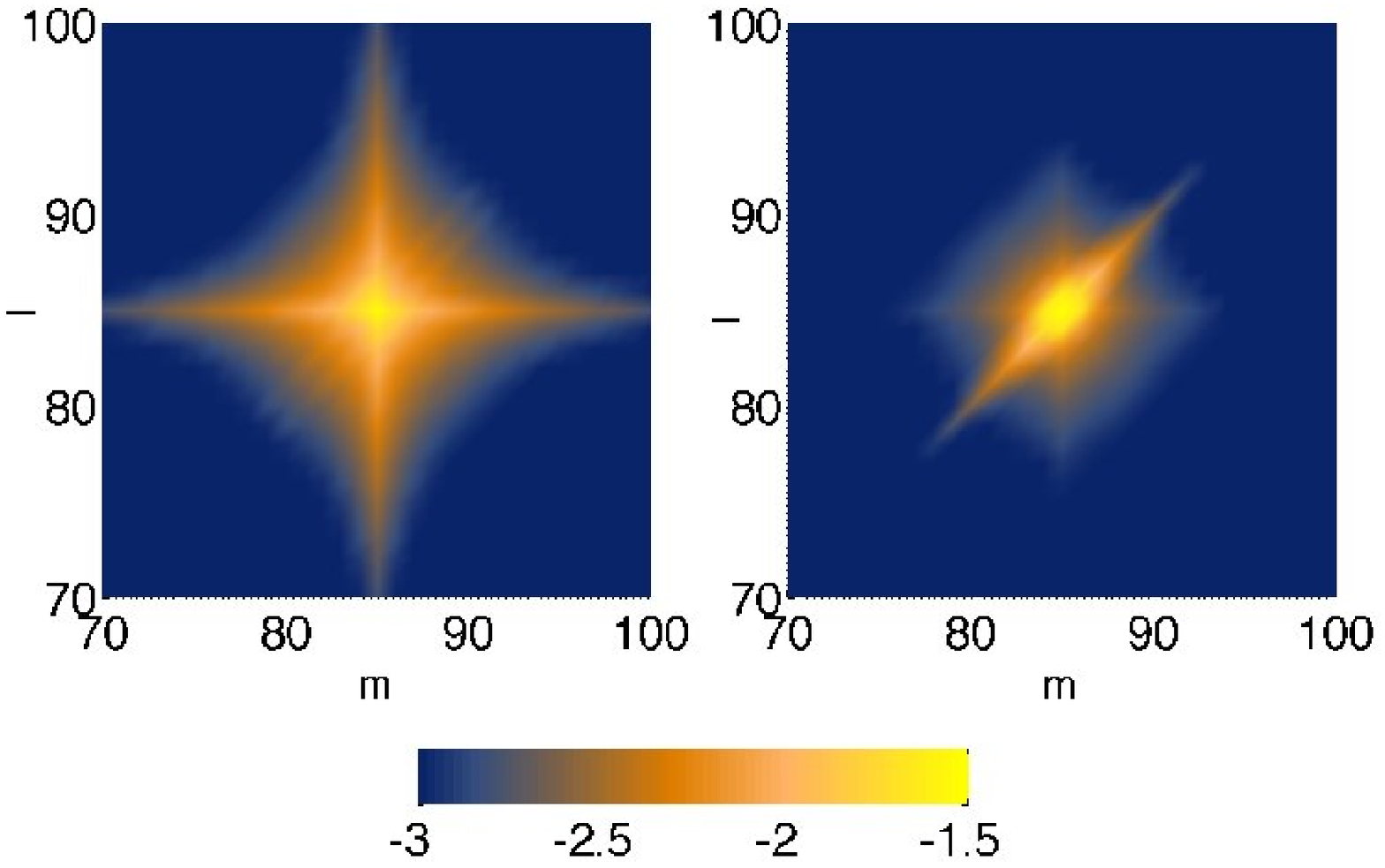}
\caption{$\log(|c_{l,m}|^2)$ versus $l$ and $m$ averaged over
time and 5000 disorder realizations.
%The
%calculations are performed for $l\le m$ corresponding to the lower
%triangle.
For better visualization results are unfolded from the irreducible
triangle shaped state space onto a square with $c_{l,m}=c_{m,l}$ for  $m\le
l$. The strength of disorder $W=2.5$ and the interaction constants
$U=0$ (a) and $U=2$ (b). Particles are initially located on the
same site at the center of a chain with $N=170$ sites.
%The second
%moments $m_2$ are $303$ and $771$, respectively, for the (a) and
%(b) cases, although from the figures one can get a feeling that
%the wave packet spreads stronger in the noninteracting case.
}
\label{fig_prb_2part}
\end{figure}
%

%
%%%%%%%%%%%%%%%%%%%%%%%%%%%%%%%%%%%%%%%%%%%%
\section{Numerical technique}
%%%%%%%%%%%%%%%%%%%%%%%%%%%%%%%%%%%%%%%%%%%%
%

We estimate the largest average localization length $\xi_2$ of the
probability density function $p_l\sim e^{-2l/\xi_2}$  [see
Eq.~(\ref{eq_pl})] using the following procedure (the
prefactor 2 in the exponent takes care of the fact that densitites instead
of wave functions are fitted).
For a
given realization we solve the eigenvalue problem and choose
only those modes ${\cal L}_{l,m}^{(q)}$ which satisfy to
the following selection rules:
\begin{itemize}
\item the center of masses
\begin{equation}
\bar l_q=\sum_{m, l\le m}^N l {\cal L}_{l,m}^{(q)2},\,\,\,\bar m_q=\sum_{m, l\le m}^N m {\cal L}_{l,m}^{(q)2}
\end{equation}
satisfy to the inequalities $|\bar l_q-N/2| \le \xi_1\,\,\,|\bar
m_q-N/2| \le \xi_1$ ($\zeta$ is of the order of the corresponding
average localization length for a single particle problem). Thus,
we take into account only those modes for which the two particles
reside in the same localization volume;
%are interacting with
%each other in a nonexponentially weak way and live far away from
%the boundaries;
\item the eigenvalues are near the bandwidth
center. We assume that similar to the case of a single particle
problem the most extended modes are with $\lambda_q \approx 0$;
\item we project ${\cal L}_{l,m}^{(q)}$ onto the modes of the
one-particle problem, calculate the amplitudes $\phi_{\mu\nu}$ in
accordance with Eq.~(\ref{eq_proj}) and find the mode  $|\mu_0,\nu_0 \rangle$
with the largest amplitude, $\max_{\mu_0,\nu_0}\phi_{\mu\nu}^2$. Such a method allows us to identify
the Fock state $|\mu_0,\nu_0 \rangle$ which dominates all others.
We then request that the eigenvalues $\lambda_{\mu_0}$ and
$\lambda_{\nu_0}$ are close to the bandwidth center. Thus, we
exclude possible cases when $\lambda_{q}$ is close to the band center, but
$\lambda_{\mu_0}$ and $\lambda_{\nu_0}$ are located at the two opposite band edges.
\end{itemize}
Having selected the modes ${\cal L}_{l,m}^{(q)}$, we compute their
probability density functions $p_l$  according to
Eq.~(\ref{eq_pl}) and shift them such that their new center of
mass are located at the center of a chain, $N/2$. Then, we
compute logarithms of the PDFs, $\ln(p_l)$ and perform a
statistical average of the PDFs over many disorder realizations
as $\langle p_l
\rangle=\exp[\langle\ln(p_l)\rangle]$. Finally, using a local
regression smoothing technique, we obtain smooth functional
dependencies of $\langle p_l \rangle$ and calculate the quantity
$\alpha=2\cdot |d(\ln \langle p_l
\rangle)/dl|^{-1}$.
In the limit of large $l$, $\alpha (l)$ should saturate at the
average two particle localization length $\xi_2$.

%
%%%%%%%%%%%%%%%%%%%%%%%%%%%%%%%%%%%%%%%%%%%%
\section{Numerical results}
%%%%%%%%%%%%%%%%%%%%%%%%%%%%%%%%%%%%%%%%%%%%
%
The dimension of the Hilbert space $p$ grows rapidly ($\sim N^2$)
with the size of a chain, so that the maximal reachable size used
in numerical computations, $N_{max}=234$. Thus, we inevitably face
finite size effects for weak disorder.
We start with the
noninteracting case $U=0$ for which $\xi_2$ must be exactly equal
to $\xi_1$. We estimate the minimal value for the strength of
disorder, respectively, maximal localization length,  $\xi_2$, at
which an error (caused by finite-size effects) is less than $10\%$
(which is the maximal error we admit). We assume that
this error depends only on the magnitude of $\xi_2$ but not on the
interaction strength $U$. Thus, the largest tolerable values for
$\xi_2$ found for the noninteracting case are also assumed to be
the limiting values for the interacting case.
For $U=0$ the lower curve in Fig.~\ref{fig_W2p0_joint}(a)
presents a smooth dependence of $\langle p_l \rangle$ on $l$.
The corresponding quantity $\alpha$ (lower curve in Fig.~\ref{fig_W2p0_joint}(b))
saturates at large distances. The obtained localization length $\xi_2$ is finally
shown in Fig.~\ref{fig_W2p0_joint}(c) and agrees well with the theoretical prediction,
however systematic deviations accumulate for weak disorder.
A recalculation of the same quantities for $U=0.2$ in  Fig.~\ref{fig_W2p0_joint}(a,b)
shows that the method appears to be applicable to the interacting case as well.
Finite size effects blurr our results substantially if $\xi_2 > 40$.

Let us discuss our results for nonzero interaction.
The ratio
$\xi_2/\xi_1$ grows with increasing interaction constant $U$, as shown for different values
of $W$ in Fig.~\ref{fig_xi2_vs_U}.
This growth is stronger, the weaker the disorder strength is. For our data,
the ratio did not substantially exceed the value 2. However, it seems plausible
that for $W < 2$ (which is not treatable with our current technique),
stronger enhancement effects could be observed.

The central result is plotted in
Fig.~\ref{fig_xi2_xi0}. Here we plot $\xi_2$ versus $\xi_1$
on log-log scales. We try to fit data for a fixed value of $U$
and different values of $W$ using power law estimates.
Both $\xi_1$ and $\xi_2$ vary less than an order of magnitude,
while a safe power law fit needs at least two orders of magnitude variations
on each variable. Nevertheless we bound the obtained variations with
two lines $\xi_2 \sim \xi_1^{1.3}$ and $\xi_2 \sim \xi_1^{1.4}$.
Such a scaling is much weaker than the any of the above predicted power laws.
It is possile that we observe the onset of the logarithmic scaling obtained
from perturbation theory (\ref{ourscaling}).

\subsection{Averaged evolution of two particles}

In order to visualize the effect of interaction on the localization of
two particles, we solve the time dependent Schr\"odinger
equation $i\partial_t |\Psi(t)\rangle={\cal \hat H}
|\Psi(t)\rangle$.  We expand $|\Psi(t)\rangle$ in terms of the
orthonormal states $|l,m \rangle$ ($l\leq m$) as
\begin{equation}
|\Psi(t)\rangle=\sum_{m,l\le m}^N  c_{l,m}(t)  |l,m \rangle.
\label{eq_expan}
\end{equation}
where the coefficients
${\cal L}_{l,m}^{(q)}$
\begin{equation}
c_{l,m}(t)=\sum_{q=1}^p \varphi_q{\cal L}_{l,m}^{(q)}
e^{-i\lambda_q t}\;. \label{eq_eigenval}
\end{equation}
Here $\varphi_q$ are the amplitudes of NMs related with the
initial amplitudes $c_{l,m}(0)=\langle l,m|\Psi(0)\rangle$ of the
two-particle states as
\begin{equation}
\varphi_q=\sum_{m,l\le m}^N c_{l,m}(0)  {\cal L}_{l,m}^{(q)}.
\label{eq_init}
\end{equation}

We launch two particles on the
same site, $l_0=m_0$, or adjacent sites, $l_0=m_0-1$, such that the
initial amplitude $c_{l,m}(0)=\delta_{l,l_0}\delta_{m,m_0}$.  We
calculate then the averaged in time square amplitude $\langle
|c_{l,m}|^2\rangle_t$  [see Eqs.~(\ref{eq_eigenval},\ref{eq_init})],
which is given by
\begin{equation}
\langle |c_{l,m}|^2\rangle_t\equiv \lim_{T\rightarrow
\infty}\dfrac{\int_0^T |c_{l,m}|^2}{T}dt=\sum_{q=1}^p
|\varphi_q|^2 {\cal L}_{l,m}^{(q)2}. \label{eq_cf_2_aver}
\end{equation}
We further average $\langle |c_{l,m}|^2\rangle_t$ over 5000
disorder realizations. In addition we perform an averaging with
respect to initial conditions, by keeping the same disorder
potential, and taking different neighboring sites as an initial
location of the particles.  Finally, we compute  the average
probability density function $\langle p_l \rangle$ using
$p_l=\dfrac{1}{2}\left(\sum_{k, l\le k}^N |c_{lk}|^2+\sum_{m, l\ge
m}^N |c_{ml}|^2\right)$.

Note that the averaged in time two-particle wavefunction $|c_{lm}|^2$
for a single disorder realization has many spots at different locations due
to resonances.
This feature is smeared out, once the averaging with respect to
disorder realizations is performed as is seen in
Fig.~\ref{fig_prb_2part}(a),(b). For the noninteracting case the obtained
distribution is elongated along the main axes. This happens because the two
particles are not correlated, and it is much more probable for them to occupy
different space regions.
However for $U=2$ the distribution is elongated along the diagonal.
This implies that the two particles are exploring more states when
being close to each other.

%We also estimate the enchancement factor of the average
%localization volume occupied by two particles, $L_2$ in comparison
%to the single particle localization volume $L_1$ (see \cite{kf10}
%for definition of $L_1$). We take $L_2=\sqrt{12m_2}$ as a
%measure of the localization volume occupied by two particles ,
%where $m_2$ is the second moment defined as $m_2 = \sum_l
%(l-\langle l \rangle)^2 p_l,\,\,\, \text{with}\,\,\,\,\langle l
%\rangle=\sum_l l  p_l$. $L_2$ turned out to scale with $L_1$
%similar to as $\xi_2$ with $\xi_1$ does.

\section{Summary} In summary, we
discussed the possible regimes of two interacting particles in a random potential.
The most interesting case of a weak Fock space disorder regime was analyzed,
and scaling laws were discussed. These results, as well as the numerical data presented as well,
show that the localization length enhancement effect is much weaker than previously assumed.
Further numerical studies are needed in order to substantiate these results. However
the current techniques are not of use for weaker disorder strength.
Therefore new computational approaches are needed in order to reach disorder
values as low as $W=0.1$, which may be enough to test the predicted weak logarithmic
scaling.

\section{Acknowledgement}The authors wish to thank I. Aleiner and B.L. Altshuler for insightful discussions.

{}


\begin{thebibliography}{99}
%
\bibitem{PWA58} P. W. Anderson, Phys. Rev. {\bf 109}, 1492 (1958).
\bibitem{KRAMER}B. Kramer and A. MacKinnon, Rep. Prog. Phys. \textbf{56}, 1469 (1993).
\bibitem{Shep_94} D.L. Shepelyansky, Phys. Rev. Lett. {\bf 73}, 2607 (1994).
\bibitem{Imry_95}Y. Imry, Europhys. Lett. {\bf 30}, 405 (1995).
\bibitem{frahm95}K. Frahm, A. M\"uller-Groeling, J.-L. Pichard, D. Weinmann, Europhys. Lett. {\bf 31}, 169 (1995).
\bibitem{Oppen_96}F. von Oppen, T. Wettig, and Jochen M\"uller Phys. Rev. Lett. {\bf 76}, 491 (1996).
\bibitem{Screib_97}R.A. R\"omer, M. Schreiber, Phys. Rev. Lett. {\bf 78}, 515 (1997).
\bibitem{Waintal99}S. De Toro Arias, X. Waintal, J.-L. Pichard, Eur. Phys. J. B {\bf 10}, 149 (1999)
\bibitem{kf10}D.O. Krimer, S. Flach,  Phys. Rev. E {\bf 82}, 046221 (2010).
\bibitem{wtl06} K. Winkler et al, Nature {\bf 441}, 853 (2006).
\bibitem{eilbeck} A.C. Scott, J.C. Eilbeck, H. Gilhoj, Physica D, {\bf 78},
194 (1994)
\bibitem{rsv99} R. A. R\"omer, M. Schreiber and T. Vojta, phys. stat. sol. {\bf 211}, 681 (1999).
%
\end{thebibliography}
\end{document}